\documentclass{acm_proc_article-sp}
\usepackage{url}
\begin{document}

\title{Assessing the Quality of Web Content}
% \subtitle{[Extended Abstract]
% \titlenote{A full version of this paper is available as
% \textit{Author's Guide to Preparing ACM SIG Proceedings Using
% \LaTeX$2_\epsilon$\ and BibTeX} at
% \texttt{www.acm.org/eaddress.htm}}}

\numberofauthors{4} %  in this sample file, there are a *total*
\author{
% 1st. author
\alignauthor
Elisabeth Lex\\
       \affaddr{Know-Center GmbH}\\
       \affaddr{Inffeldgasse 21a}\\
       \affaddr{Graz, Austria}\\
       \email{elex@know-center.at}
\alignauthor
% 2nd. author
Inayat Khan\\
       \affaddr{Institute for Computer Graphics and Vision Graz}\\ 
       \affaddr{Graz University of Technology }\\
       \affaddr{Inffeldgasse 16 }\\
       \affaddr{Graz, Austria}\\
       \email{khan@icg.tugraz.at }
\alignauthor
Horst Bischof\\
       \affaddr{Institute for Computer Graphics and Vision Graz}\\ 
       \affaddr{Graz University of Technology }\\
       \affaddr{Inffeldgasse 16 }\\
       \affaddr{Graz, Austria}\\
       \email{bischof@icg.tugraz.at }
\and
\alignauthor
Michael Granitzer\\
       \affaddr{Know-Center GmbH}\\ 
       \affaddr{Knowledge Management Institute}\\ 
       \affaddr{Graz University of Technology}\\
       \affaddr{Inffeldgasse 21a}\\
       \affaddr{Graz, Austria}\\
       \email{mgrani@know-center.at}
% 3rd. author
% \alignauthor Lars Th{\o}rv{\"a}ld\titlenote{This author is the
% one who did all the really hard work.}\\
%        \affaddr{The Th{\o}rv{\"a}ld Group}\\
%        \affaddr{1 Th{\o}rv{\"a}ld Circle}\\
%        \affaddr{Hekla, Iceland}\\
%        \email{larst@affiliation.org}
% \and  % use '\and' if you need 'another row' of author names
% % 4th. author
% \alignauthor Lawrence P. Leipuner\\
%        \affaddr{Brookhaven Laboratories}\\
%        \affaddr{Brookhaven National Lab}\\
%        \affaddr{P.O. Box 5000}\\
%        \email{lleipuner@researchlabs.org}
% % 5th. author
% \alignauthor Sean Fogarty\\
%        \affaddr{NASA Ames Research Center}\\
%        \affaddr{Moffett Field}\\
%        \affaddr{California 94035}\\
%        \email{fogartys@amesres.org}
% % 6th. author
% \alignauthor Charles Palmer\\
%        \affaddr{Palmer Research Laboratories}\\
%        \affaddr{8600 Datapoint Drive}\\
%        \affaddr{San Antonio, Texas 78229}\\
%        \email{cpalmer@prl.com}
}
% There's nothing stopping you putting the seventh, eighth, etc.
% author on the opening page (as the 'third row') but we ask,
% for aesthetic reasons that you place these 'additional authors'
% in the \additional authors block, viz.
% \additionalauthors{Additional authors: John Smith (The Th{\o}rv{\"a}ld Group,
% email: {\texttt{jsmith@affiliation.org}}) and Julius P.~Kumquat
% (The Kumquat Consortium, email: {\texttt{jpkumquat@consortium.net}}).}
% \date{30 July 1999}
% Just remember to make sure that the TOTAL number of authors
% is the number that will appear on the first page PLUS the
% number that will appear in the \additionalauthors section.
\permission{Copyright is held by authors/owners.}
\conferenceinfo{ECML/PKDD Discovery Challenge,}{September 20th,2010.}
\copyrightetc{Barcelona, Spain}

\maketitle
\begin{abstract}
This paper describes our approach towards the
ECML/PKDD Discovery Challenge 2010. The challenge consists of three tasks: (1) a Web genre and facet classification task for English hosts, (2) an English quality task, and (3) a multilingual quality task (German and French). In our approach, we create an ensemble of three
classifiers to predict unseen Web hosts whereas each classifier is trained on a different feature set. Our final NDCG on the whole test set is $0.537$ for Task 1, $0.844$ for Task 2, and $0.823$ (French) and $0.793$ (German) for Task 3, which ranks fourth place in the ECML/PKDD Discovery
Challenge 2010.
% take
% content based features as well as link based features into account. Our main focus is to exploit content based data, however, especially for
% the Web Spam detection task, link based features contribute well. To address the
% quality problem, the organizers of the ECML/PKDD Discovery Challenge 2010 divided
% the problem into three tasks: a Web genre detection task (Classification task
% (English)) , an English quality task, and a multilingual Quality task (German and
% French). Quality is measured as an aggregate function of genre, trust, factuality
% and bias and spam has lowest (0) quality.
\end{abstract}

%A category including the fourth, optional field follows...



\section{Introduction}
On the Web, a huge amount of information and content is available. However, this content drastically varies from high quality to abusive content and spam~\cite{Agichtein2008}. From a Web archive point of view, the usefulness of content obtained from web crawls is sometimes questionable, especially in respect to information quality. If quality measures or rankings would be available in addition to the content itself, the archival would be improved as it can be automatically decided whether it is worth to archive a particular Web content or not.
The ECML/PKDD Discovery Challenge 2010 aims at developing automatic methods to estimate the overall rank, quality, and importance of Web content\footnote{\url{http://www.ecmlpkdd2010.org/articles-mostra-2041-eng-discovery_challenge_2010.htm}}. The goal is to support organizations to prioritize the gathering, storing and organization of Web pages.
 
\section{Tasks}
\label{s:tasks}
The challenge consists of three tasks: (i) a classification task to assess the Web genre and information quality facets like neutrality, bias, and trustiness, (ii) an English quality task whereas the quality of a Web site is measured as an aggregate function of its genre and its neutrality, bias and trustiness, and (iii) a multilingual quality task where the quality of German and French Web sites has to be assessed. 

\subsection{Task 1}
\label{s:task1}
The goal of Task 1 is to classify English Web hosts into a set of categories: Web
Spam, News/ Editorial, Commercial, Educational/Research, Discussion,
Personal/Leisure, and to assess the level of neutrality, bias, and trustiness on
a scale from 1 to 3 whereas 3 denotes normal and 1 problematic content. The
result of Task 1 is a ranked list whereas we rank the test hosts by classifier
confidence.

\subsection{Task 2}
\label{s:task2}
The aim of Task 2 is to measure the quality of the English Web hosts whereas the
quality is determined as an aggregate function of the host's genre, its
neutrality, bias, and trustiness. The facets neutrality, bias, and trustiness
cover the intrinsic content quality, as described by Huang et al.
in~\cite{Huang1999}. The overall quality score is derived by combining the
results retrieved in Task 1 according to the following rule:
\begin{verbatim}
utilityScore = 0;
if (News-Edit OR Educational) {
	value = 5;
} else if (Discussion) {
	value = 4;
} else if (Commercial OR Personal-Leisure) {
	value = 3;
}
if (neutrality == 3) value += 2;
if (bias == 1) value -= 2;
if (trustworthiness == 3) value +=2;
\end{verbatim}

The rationale behind this definition of quality is that the challenge organizers
define quality with regard to the needs of an Internet archive. Therefore, the
categories News and Educational have the highest quality.
Also, the rule implies that quality content should exhibit trust, no bias, and
neutrality. Consequently, Web Spam hosts have by default the lowest quality. The
result of Task 2 is also a list ranked by classifier confidence.

\subsection{Task 3}
\label{s:task3}
Task 3 aims at assessing the quality of German and French Web hosts since in the
.eu domain, a lot of content is available in other languages than English. The
focus in this task is on two major European languages, German and French. The
quality of the German and French hosts is also derived using the above rule and
as a result, a list ranked by classifier confidence is obtained.

\section{Dataset and Features}
\label{s:dataset}

The dataset for the Discovery Challenge 2010 is based on a crawl of
the .eu domain provided by the European Archive
Foundation\footnote{\url{http://datamining.sztaki.hu/?q=en/DiscoveryChallenge/}}. 
The dataset contains a collection of annotated Web hosts labeled by the
Hungarian Academy of Sciences (English), European Archive Foundation (French) and L3S
Hannover (German)~\cite{Benczur2010}. Table 1 shows the number of English training samples for each class: the dataset is in most cases highly imbalanced towards the positive class. Note that while the genre categories are mutually exclusive, the quality categories are not. 

 \begin{table*} [!htbp]
 \centering
 \caption{Number of training samples}
 \begin{tabular}{|l|c|c|} \hline
Category & Positive Samples [\%] & Negative Samples [\%] \\ \hline
WebSpam & 4 & 96\\ \hline % 4 vs 96
News/Editorial & 4.7 & 95.3 \\ \hline % 4.7% 95.3
Educational/Research & 43 & 57 \\ \hline %43, 57
Personal/Leisure &  23.7 & 76.3 \\ \hline % 23.7 
Commercial &  45.4  & 54.6 \\ \hline %45.4
Discussion &  5.3 & 94.7\\ \hline %5.3
Bias & 1.7 &  98.3 \\ \hline %1.7
Neutrality & 96.6 & 3.4 \\ \hline %96.6
Trustworthiness & 98.1 & 1.9\\ \hline %
\end{tabular}
 \end{table*}
  
\subsection{Features}
\label{s:features}
In the dataset, different types of features are provided. Most features were assessed on a per host level, only the natural language processing features are available on a large set of sample pages. The features are described in more detail in the next paragraphs. Note that as dataset backend, we created feature vectors for each feature set which we then stored in an Apache Lucene~\footnote{http://lucene.apache.org/} index. This resulted in an index size of approximately 19 GB.

%Also, we considered the hostnames because we noticed that a lot of offensive
%content like adult can be derived from analysing only the names of the domains.
%This would clearly be feasible if the task is to assess adult content, however,
%in the training set, too less of such hosts were given, even tough the test set
%contained a lot.

\subsubsection{Link Features}
\label{s:linkfeatures}

The provided link based features were derived from the Web graph and are
available on a per host level. The feature set contains features like the
in-degree, the out-degree, the PageRank, the edge reciprocity, the
assortativity coefficient, and the TrustRank, summing up to 176 features.

\subsubsection{Content based Features}
\label{s:contentfeatures}

The content based features are also available on a per host level. This
feature set contains features like the number of words in the homepage or the average length of the title. They were proposed in~\cite{Castillo2007} to detect Web spam based on content. In our setting, we exploited all given content based features (95 features).

\subsubsection{Natural Language Processing Features}
\label{s:nlpfeatures}

The Natural Language Processing (NLP) features are available per URL in contrast
to the other feature sets. They were processed by the LivingKnowledge
project\footnote{http://livingknowledge-project.eu/}. Included in this feature
set are the counts for sentence, token, character, the count of various
Part-of-Speech (POS) tags, etc. Therefore, these features cover style based properties. Generally,
stylometric features are well suited for assessing quality facets like
neutrality since they are inherently topic independent~\cite{Lex2010a, Lex2010}.
We used all NLP features except the most common bigrams - since they were often
null, resulting in 180 NLP features.

\subsubsection{Term Frequencies}
\label{s:tf}
This feature set consists of the host level aggregate term vectors of the most
frequent terms. Note that the top 50,000 terms are considered after eliminating
stop words. The term frequency is computed over an entire host while the
document frequency is on page level. We exploit the term frequency and the document
frequency to weight the features by tf-idf.

\section{Approach}
\label{s:approach}

In our approach, we implemented an ensemble classifier strategy to exploit all types of features that
were provided for the challenge. We addressed each classification task as a binary
classification strategy. More specifically, we classified the test hosts
into the positive versus the negative class using the different classifiers. We then combined the classification results based
on a majority voting whereas we assigned the test hosts to the winner with the
maximum classifier confidence.
 
For the multi language quality task (Task 3), we considered only the link based and content based features derived from the English training hosts. The training set for both the German and French hosts contains only a few annotated hosts. Therefore, we exploited the link based features for the multilingual quality task since they are inherently language independent. Also, we considered the content based features since originally, they were proposed by Castillo et al~\cite{Castillo2007} to detect spam. Since spam is typically not identified by language, our assumption was that the content based features can also be exploited over different languages. 

% \section{Algorithms}
% \label{s:algorithms}
% 
% We computed the mutual information to extract features that 
% contribute most to a label. For instance, for the category Spam versus Non Spam, especially the 
% ratio of trustrank to pagerank is a discriminate and good feature.

\subsection{Classifiers}
\label{s:classifiers}

For our ensemble based approach, we used three different
classification algorithms. Firstly, we exploited the implementation of a J48
decision tree given in Weka~\cite{hall2009} whereas we set $C= 0.25$ and $M=2$. To compensate the imbalance in the
category representation in the given training set, we applied a filter based on
Synthetic Minority Oversampling Technique (SMOTE)~\cite{Chawla2002}. In the
SMOTE technique, artificial training samples are generated for the minority
class based on the k nearest neighbours of a training item. Therefore, the
minority class is oversampled exploiting the articial training samples. It is also worth mentioning that we also
applied random sampling at first, however, SMOTE gives much better results. Note that we used the SMOTE
implementation given in Weka~\cite{hall2009}. We set the number of nearest neighbours to 5, the percentage to 100, and
the random seed to 1. Additionally, we normalized the feature values with a normalization filter from Weka. 

Secondly, we applied a centroid based classifier, the Class-Feature-Centroid
Classifier (CFC)~\cite{guan09} which is known to outperform Support Vector
Machines in certain settings. The CFC implements a highly discriminative term weighting scheme based on the inter term distribution and the intra term distribution. We already successfully used the CFC classifier for genre classification in English blogs~\cite{Lex2010b}.

Thirdly, we applied a Support Vector Machine (SVM) based on LibLinear~\cite{fan2008} since SVMs are among the best text classification
algorithms and especially the LibLinear is known to be fast and efficient.

In our approach, we used these three classifiers with different feature sets: On the term frequencies, we applied the CFC algorithm since its
highly discriminant abilities serves best in this setting. The CFC algorithms needs real terms to compute its discriminative weighting scheme and
fortunately, the challenge organizers also provided a dictionary of the 50000 top terms. Therefore, we could make use of this highly performant algorithm. Especially for topic driven categories like \emph{News/Editorial} and \emph{Educational/Research} the CFC served well. 

On the link based and content based features, we applied the J48 classifier with the SMOTE filter since cross-validation experiments on the training set revealed that this classifier deals best with the imbalance problem. Note that the J48 classifier has already been successfully applied to a similar problem of spam classification, as described by Castillo et al in~\cite{Castillo2007} with the only difference that they used it as a base classifier for a cost-sensitive classifier. In our experiments, we also evaluated a cost sensitive classifier with J48 and similar parameters as described in~\cite{Castillo2007}, however the SMOTE based approach outperformed the cost sensitive classifier. 

On the natural language processing features, we worked with the LibLinear
implementation of a SVM. This decision was based on practical reasons only since
in this case, there is a large amount of feature vectors (approx. 23M) because
the natural language processing features were assessed on a page level - in contrast to all other features which were assessed on a per host level. Clearly, the LibLinear is not the best algorithm in this setting but it is very fast and highly performant. To determine the best performing cost parameter C, we conducted a grid search and identified $C=0.04$ as best.
% 
% Unbalanced data: First, we tried to apply a cost-sensitive classifier with a
% J48 decision tree as recommended by Castillo et al in~\cite{Castillo2007}. We
% varied the parameters for the cost matrix. In a 10-fold crossvalidation on the
% training data, especially for the Spam category, the results were okay.
% However, on the test set, a SMOTE based approach outperformed the
% cost-sensitive classifier.

\section{Results}
\label{s:results}

%In this section, the tasks are described briefly and the results achieved with
%our methods are outlined.
%All results are derived following a 10-fold crossvalidation strategy. Note that
%the results are given in f-measure.

The results for Task 1 are given in Table 1. Note that the evaluation is conducted in terms of the evaluation metric Normalized
Discounted Cumulated Gain (NDCG)~\cite{Jaervelin2000}.
The results for Task 1 reveal that the category Educational achieves the best results in terms of NDCG. We manually examined a number of test hosts and identified that the categories News/Editorial and Educational/Research are quite hard to separated with the given features. A reason for this might be that both categories exhibit a similar writing style (factual, neutral, rather long and complex words) which results in similar content based and natural language processing features. Also, over both categories, similar terms are used. 
 \begin{table} [!htbp]
 \centering
 \caption{Results for Task 1}
 \begin{tabular}{|l|c|} \hline
Category & NDCG \\ \hline
WebSpam & 0.473 \\ \hline
News/Editorial & 0.416 \\ \hline
Commercial & 0.694 \\ \hline
Educational/Research & 0.688 \\ \hline
Discussion & 0.531 \\ \hline
Personal/Leisure & 0.583 \\ \hline
Trustiness & 0.397 \\ \hline
Bias &  0.540 \\ \hline
Neutrality & 0.51\\ \hline
Average & 0.537\\ \hline
\end{tabular}
 \end{table}

To improve the results, we tried to correct misclassifications in the category
\emph{News}. We applied a binary classification on the hosts the classifier ensemble predicted as News. In contrast to the earlier experiments, we performed not binary decisions between the positive and the negative class (News versus Non News) but binary decisions between News and each other category. For example, we evaluated News versus Web Spam.
We introduced the following rule: if more than two sub classifiers assigned the 
test host to a non news category, we multiplied the original classifier confidence with a factor of $0.4$ to lower the confidence of the first prediction. 
However, if we compare the results achieved for News from the first submission (0.442) with the second submission (0.416), the described post processing actually reduces the NDCG ranking. A possible explanation can be that with the post processing of solely one category, the overall ranking for the category changes too much. This is something we have to investigate in more detail.

The results derived from Task 1 are then directly used to compute the quality of the English test hosts and further to rank the English hosts by their quality. The results for Task 2 are shown in Table 3:
 \begin{table} [!htbp]
 \centering
 \caption{Results for Task 2}
 \begin{tabular}{|l|c|} \hline
Language  & NDCG \\ \hline
English & 0.844\\ \hline\end{tabular}
 \end{table}
As one can see, the quality of the English hosts can be assessed quite well. This is clearly due to the fact that in Task 1, we were able to assign the category Educational/Research with a rather good confidence, since this category has a high influence in the final quality function.
 \begin{table} [!htbp]
 \centering
 \caption{Results for Task 3}
 \begin{tabular}{|c|c|} \hline
Language  & NDCG \\ \hline
German & 0.792\\ \hline
French & 0.823 \\ \hline \end{tabular}
 \end{table}
In the multilingual setting for Task 3, we achieve good results for the French
and German hosts, even though we used only the link based and content based features derived from the English training hosts in this case. This reveals that these features are rather language independent, at least for indoeuropean languages, and robust. The results for Task 3 are shown in Table 4.

\section{Conclusions and Future Work}
\label{s:conclusion}
In our approach towards the ECML/PKDD Discovery Challenge 2010, we exploited all
provided features in an ensemble classifier setting. We applied three different
classifiers whereas each classifier was trained on a different feature set. As a
result, our approach ranks fourth overall in the challenge. Our experiments
reveal that even if the NDCG is low for some categories like Web Spam, News/Editorial, and Bias, the quality of the Web hosts can be assessed with a high NDCG of $0.844$ in the monolingual setting (English hosts), and a NDCG of $0.793$ (German) and $0.823$ (French) in the multilingual setting. For future work, we intend to focus on feature selection to extract the best features for each task. First experiments with mutual information revealed that for instance the ratio TrustRank to PageRank is a very good feature to distinguish low quality content from regular content. Besides, we work on an extension of the genre classification with a cross modal strategy where we add an image classifier to our ensemble. First experiments revealed that at least for the categories personal and commercial, adding an image classifier contributes well to the ensemble decision to predict the host's genre.

%ACKNOWLEDGMENTS are optional
\section{Acknowledgments}
The Know-Center GmbH Graz is funded within the Austrian COMET Program -
Competence Centers for Excellent Technologies - under the auspices of the
Austrian Ministry of Transport, Innovation and Technology, the Austrian Ministry
of Economics and Labor and by the State of Styria. COMET is managed by the
Austrian Research Promotion Agency FFG.
\bibliographystyle{abbrv}
\bibliography{sigproc}  % sigproc.bib is the name of the Bibliography in this case
% ACM needs 'a single self-contained file'!
%\balancecolumns

\balancecolumns
% That's all folks!
\end{document}